\documentclass[12pt]{article}

\usepackage{graphicx}
\usepackage{amsmath}
\usepackage{a4wide}
\usepackage{psfrag}
\usepackage{bbm}
\usepackage{natbib}

\newcommand{\ve}{\boldsymbol}
\bibpunct[,]{(}{)}{,}{a}{}{,}

\psfrag{a}{(a)}
\psfrag{b}{(b)}
\psfrag{c}{(c)}
\psfrag{xi}{$\xi$}
\psfrag{xistrich}{$\xi'$}
\psfrag{xistrich2}{$\xi'_{2}$}
\psfrag{R}{$R$}
\psfrag{T}{$T$}
\psfrag{T2}{$T_{2}$}
\psfrag{deka1}{1.6 nm d-phase}
\psfrag{deka2}{1.2 nm d-phase}
\psfrag{D}{D}
\psfrag{U}{U}
\psfrag{N}{N}
\psfrag{H}{H}
\psfrag{P}{P}
\psfrag{D2}{D'}
\psfrag{S2}{S'}
\psfrag{B2}{B'}
\psfrag{H2}{H'}
\psfrag{AR}{AR}
\psfrag{AR2}{AR'}
\psfrag{OR}{OR}
\psfrag{OR2}{OR'}

\begin{document}

\title{Tiling models for metadislocations in AlPdMn approximants}

\author{M.~Engel\footnote{Author for correspondence. Email:
  mengel@itap.physik.uni-stuttgart.de}~ and H.-R.~Trebin\\
  {\normalsize Institut f\"ur Theoretische und Angewandte Physik,}
  {\normalsize Universit\"at Stuttgart,} \\
  {\normalsize Pfaffenwaldring 57, D-70550 Stuttgart, Germany}\\
}

\maketitle

\begin{abstract}

The AlPdMn quasicrystal approximants $\xi$, $\xi'$, and $\xi'_{n}$ of the
1.6~nm decagonal phase and $R$, $T$, and $T_{n}$ of the 1.2~nm decagonal phase
can be viewed as arrangements of cluster columns on two-dimensional
tilings. We substitute the tiles by Penrose rhombs and show, that alternative
tilings can be constructed by a simple cut and projection formalism in three
dimensional hyperspace. It follows that in the approximants there is a
phasonic degree of freedom, whose excitation results in the reshuffling of the
clusters. We apply the tiling model for metadislocations, which are special
textures of partial dislocations.

\end{abstract}

\section{Introduction}

A quasiperiodic structure can be described as a cut through a hyperlattice
decorated with atomic surfaces. As a consequence the atoms are arranged in a
finite number of different local environments frequently leading to a
substructure of highly symmetric clusters. The cluster positions can again be
modelled by a simpler decoration consisting in the simplest case of exactly
one atomic surface (for the cluster centre) per hyperlattice unit cell. The
displacement of the cut space (phasonic displacement) is a discrete degree of
freedom, called phasonic degree of freedom. It can be excited locally, leading
to a rearrangement of the clusters by correlated atomic jumps. This view is
supported by recent diffraction data of coherent phason modes in
i(cosahedral)-AlPdMn \cite[]{itapdb:Coddens1999, itapdb:Francoual2003} and by
in situ observations of phason jumps via high-resolution transmission electron
microscopy (HRTEM) in d(ecagonal)-AlCuCo \cite[]{itapdb:Edagawa2002}.

The cut formalism cannot be applied directly if there is a gradient in
the phasonic displacement. This is examplified by the extreme case of a cut
space running inbetween all atomic surfaces and not touching any of
them. Hence we resort to another method: We substitute the atomic surfaces by
atomic hypervolumes \cite[]{itapdb:Engel2005} of the same dimension as the
hyperspace. For the construction two different spaces are needed: Those atomic
hypervolumes that are cut by the (possibly deformed) cut space
$E_{\text{Cut}}$ are selected. The centre of each selected atomic hypervolume
is projected onto the projection space $E_{\text{Proj}}$. By shearing the cut
space, i.e. by introducing a linear phasonic displacement, periodic
approximants are created. Phasonic degrees of freedom also can exist in these
and play a fundamental role for phase transitions
\cite[]{itapdb:Edagawa2004}.

Here we discuss linear defects, metadislocations, in the phasonic degree of
freedom for approximants of the AlPdMn system. This system is especially
adequate for the examination of phasonic degrees of freedom since a stable
i-phase, a stable 1.2~nm d-phase, a metastable 1.6~nm d-phase (which is
assumed to be a solid solution of Mn in d-AlPd \cite[]{itapdb:steurer2004}),
and a large variety of approximants have been observed in the phase diagramm
\cite[]{itapdb:Klein2000a}. All of them, as well as several binary AlPd and
AlMn quasicrystals and approximants are related structurally.

\section{Tiling models}

A hyperspace model for i-AlPdMn has been proposed by
\cite{itapdb:Katz1993}. It uses a six dimensional face-centred hyperlattice
$F^{6D}$ with lattice constant $2l^{6D}=1.29$~nm and serves as a starting point
since to our knowledge all newer, more complicated hyperspace models are
refinements. It has been shown, that the approximants $\xi$ and $\xi'$
\cite[]{itapdb:Beraha1997} of the 1.6~nm d-phase and the approximants $R$ and
$T$ \cite[]{itapdb:Beraha1998} of the 1.2~nm d-phase are described on an
atomistic level by a shear in this model. However the authors had to introduce
an additional mirror symmetry to assure full tenfold symmetry. The main
building units are Mackay-type clusters, whose centres are projected from the
hyperspace by using one atomic surface per hyperlattice unit cell. It is a
subset of a triacontahedron, deflated by $\tau=\frac{1}{2}(\sqrt{5}+1)$ with
respect to the canonical triacontahedron, which is the projection of the
hypercube with edge length $l^{6D}$ on the orthogonal complement of
$E_{\text{Proj}}$. The distance of the cluster centres is the shortest
projection of neighboring $F^{6D}$-sites $\ve{e}_{i}\pm\ve{e}_{j}$ multiplied
by the deflation factor: $t^{6D}=\frac{1}{5}\sqrt{10}\sqrt{\tau+2}l^{6D}\simeq
0.78$~nm.

The relation of the approximants to the i-phase is given by two consecutive
shears of $E_{\text{Cut}}$ in the hyperspace: The first shear changes the
cluster arrangement in direction of a fivefold axis $\ve{e}_{1}$. Together
with the introduction of a mirror plane this results in a decagonal
quasicrystal. The clusters are then aligned in columns parallel to the tenfold
axis, so that the structures can be described by two-dimensional tilings, which
are the projections in the column direction. The second shear rearranges the
columns perpendicular to $\ve{e}_{1}$. We will now consider the tilings for
the 1.6~nm and the 1.2~nm phases separately.

\begin{figure}
\begin{center}
\includegraphics[height=18cm]{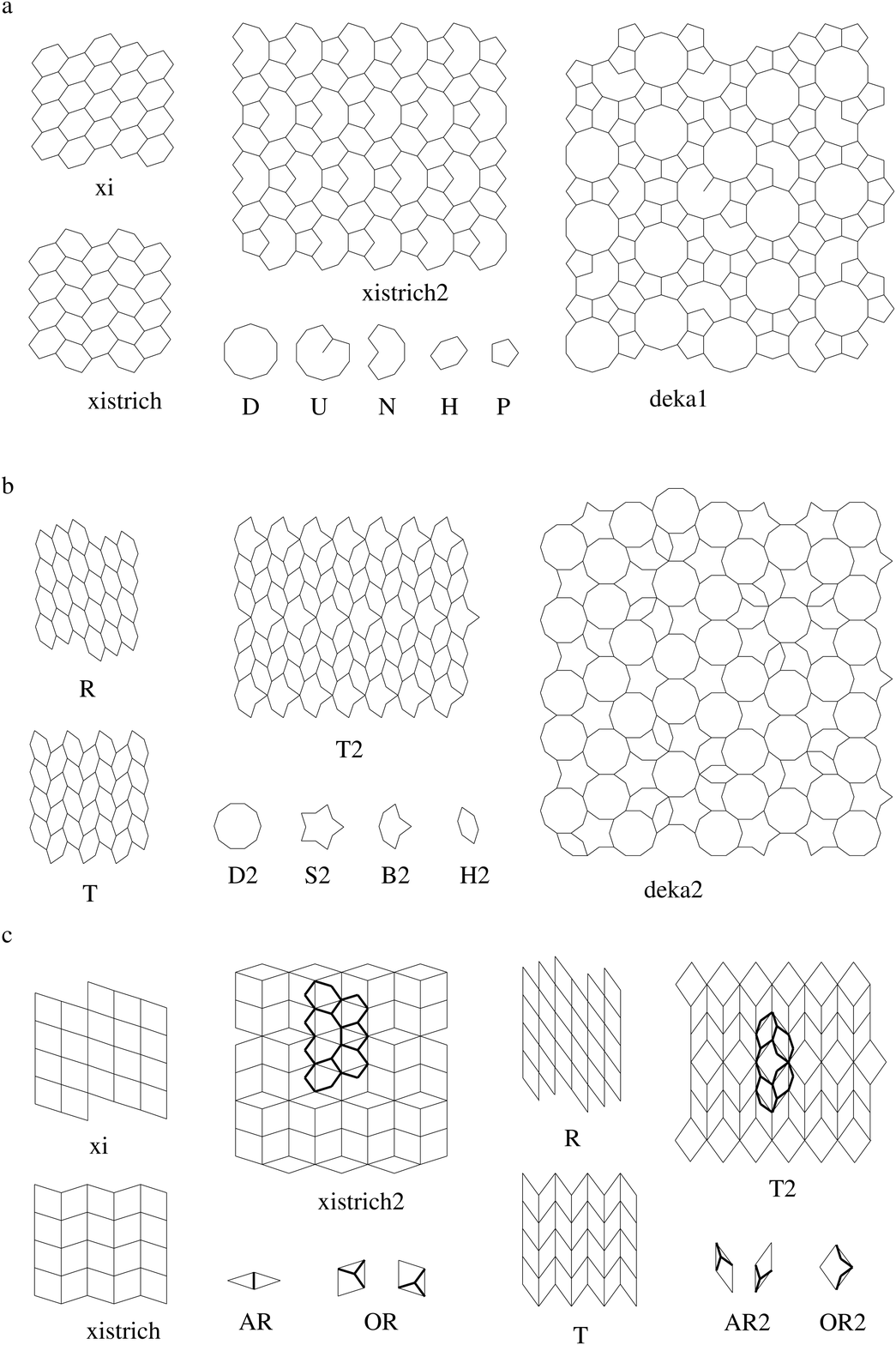}
\caption{Calculated tilings for various approximants of the AlPdMn
  i-phase. (a) 1.6 nm phases: The tiling of the d-phase is the T\"ubingen
  Triangle Tiling (TTT). (b) 1.2 nm phases: The D' centers lie on a $\tau^{2}$
  inflated TTT. (c) The tiles can be substituted with Penrose rhomb tiles. The
  substitution is different for the 1.6 nm phases and the 1.2 nm phases.
\label{fig:tilings}}
\end{center}
\end{figure}

Neighboring clusters in the 1.6~nm d-phase and its approximants lie on planes
perpendicular to $\ve{e}_{1}$. Thus the tile length is $t^{6D}$ as defined
above. The $\xi$- and the $\xi'$-phase are built from flattened hexagons (H)
arranged in parallel and in alternated orientation respectively. By
introducing additional rows of pentagons (P) and nonagons (N) between each
$n-1$ rows of alternated Hs in the $\xi'$-phase we obtain the
$\xi'_{n}$-phases. Such a PN-row is called a phason plane, which is justified
since it is elongated in the $\ve{e}_{1}$ direction. As we will see, phason
planes play an important role for the cluster reshuffling resulting from an
excitation of the phasonic degree of freedom. Further tiles, the decagon (D)
and an U-shaped tile (U) are observed in the 1.6~nm d-phase
(Fig.~\ref{fig:tilings} (a)).

Neighboring clusters in the 1.2~nm d-phase and its approximants lie on two
planes staggered perpendicular to $\ve{e}_{1}$ with distance
$\frac{1}{5}\sqrt{10}l^{6D}\simeq 0.41$~nm. The resulting tile length is
$t'^{6D}=\frac{1}{5}\sqrt{10}\tau l^{6D}\simeq 0.66$~nm. The $R$- and the
$T$-phase are built from elongated hexagons (H') arranged in parallel and in
alternated orientation respectively. The $T_{n}$-phases are created by
introducing into the $T$-phase additional rows of boat-shaped tiles (B')
between each $n$ rows of alternated H's. A B'-row is again called a
phason-plane. For the 1.2~nm d-phase additionally a decagon (D') and a
star-shaped tile (S') are needed (Fig.~\ref{fig:tilings} (b)).

By substituting the H, P, N, H', and B' tiles with acute rhombs (AR), (AR')
and obtuse rhombs (OR), (OR') as shown in Fig.~\ref{fig:tilings} (c) we obtain
new tilings for the approximants, which can be interpreted as approximants of
the Penrose-tiling. An H is substituted by an OR, while a phason plane
corresponds to a combination of an AR-row and an OR-row. So the
$\xi'_{n}$-phase has $n$ OR-rows inbetween neighboring AR-rows. Similarly the
$T_{n}$-phase has $n$ AR-rows inbetween neighboring phason planes, represented
by OR-rows. The rhombs occuring in the new tilings for the $\Xi$-approximants
($\xi$, $\xi'$, $\xi'_{n}$), as well as for the $T$-approximants ($R$, $T$,
$T_{n}$) both only need three of the five basis vectors of the Penrose-tiling
to be constructed. Therefore the tilings can be modelled in a simple
three-dimensional hyperspace with the $\mathbbm{Z}^{3}$-lattice and lattice
constant $l^{3D}=\tau\sqrt{\tau+2}l^{6D}\simeq 1.99$~nm. The projection
matrices are ($s_{i}=\sin(2\pi\frac{i}{5})$, $c_{i}=\cos(2\pi\frac{i}{5})$):
\begin{equation}
\pi^{\parallel}_{\Xi}=\frac{1}{5}\sqrt{10}\left(\begin{array}{ccc}
s_{0} & s_{1} & s_{4}\\
c_{0} & c_{1} & c_{4}\\
\end{array}\right),\qquad
\pi^{\parallel}_{T}=\frac{1}{5}\sqrt{10}\left(\begin{array}{ccc}
s_{0} & s_{2} & s_{3}\\
c_{0} & c_{2} & c_{3}\\
\end{array}\right),
\end{equation}
leading to an edge length of the tiles:
$t^{3D}=\frac{1}{5}\sqrt{10}l^{3D}\simeq 1.26$~nm. There is one atomic
hypervolume per unit cell, which is just the unit cell, and one phasonic
degree of freedom. This three-dimensional hyperspace is the simplest model for
a phasonic degree freedom besides the Fibonacci-chain.

\section{Metadislocations}

In the formalism of atomic hypervolumes a dislocation can be introduced into
a tiling by a generalised Voltera process \cite[]{itapdb:Engel2005}. It is
uniquely characterised by a translation vector of the hyperlattice, the
Burgers vector $\ve{b}$ (here: $\ve{b}^{3D}=(b_{1},b_{2},b_{3})$,
$b_{i}\in\mathbbm{Z}$), that splits up into a phononic component
$\ve{b}^{\parallel}=\pi^{\parallel}\ve{b}$ (deforming the tiles) and a
phasonic component $\ve{b}^{\perp}$ (rearranging the tiles). The latter can
only be calculated from the full six-dimensional Katz-Gratias model. If it is
not zero, such a dislocation is a partial dislocation. By (i) extending the
linear theory of elasticity to the hyperspace, (ii) approximating the phasonic
degree as continuous, and (iii) assuming isotropy in the strain fields, the
line energy $E$ of a dislocation is expressed as:
\begin{equation}\label{eq:energy}
E=c_{\text{phon}}\|\ve{b}^{\parallel}\|^2+c_{\text{phas}}\|\ve{b}^{\perp}\|^2+
c_{\text{coupl}}\|\ve{b}^{\parallel}\|\|\ve{b}^{\perp}\|.
\end{equation}

Besides a phononic contribution with material constant $c_{\text{phon}}$ and a
phasonic contribution with $c_{\text{phas}}$, a coupling term is present with
$c_{\text{coupl}}$. According to experiment we assume $c_{\text{phon}}\gg
c_{\text{phas}}\approx c_{\text{coupl}}$. Since stable dislocations are those
with the lowest energy, we have to minimise $\|\ve{b}^{\parallel}\|$. We will
discuss this in parallel for dislocations in the $\Xi$- and the
$T$-approximants. The minimization yields $b_{2}=b_{3}$ in both
cases. Furthermore we have $b_{1}=-\tau^{-1}b_{2}$ for $\ve{b}^{3D}_{\Xi}$ and
$b_{1}=\tau b_{2}$ for $\ve{b}^{3D}_{T}$. Here we approximate $\tau^{-1}$ by
the fractions $F_{m-1}/F_{m}$ and $\tau$ by the fractions $F_{m+1}/F_{m}$
respectively. $(F_{m})_{m\in\mathbbm{N}}$ are the Fibonacci numbers with start
values $F_{1}=F_{2}=1$. Finally the Burgers vectors of stable dislocations
are: $\ve{b}^{3D}_{\Xi}=(F_{m-1},-F_{m},-F_{m})$ and
$\ve{b}^{3D}_{T}=(F_{m+1},F_{m},F_{m})$. Interestingly they correspond to the
same six-dimensional Burgers vectors:
\begin{equation}\label{eq:burgers}
\ve{b}^{6D}_{\Xi}=\ve{b}^{6D}_{T}=(0,0,-F_{m-2},F_{m-1},F_{m-2},F_{m-1}).
\end{equation}
Hence it suffices to consider both cases together for the rest of our
calculations. The phononic component
$\ve{b}^{\parallel}=\ve{b}^{\parallel}_{\Xi}=\ve{b}^{\parallel}_{T}$
is perpendicular to the phason-planes (in the vertical direction in
Fig.~\ref{fig:tilings}). We get $\|\ve{b}^{\parallel}\|=\tau^{-m}t^{3D}$ and
$\|\ve{b}^{\perp}\|=\tau^{m-3}t^{3D}$. Substituting this into
(\ref{eq:energy}), we have $E=\left[c_{\text{phon}}\tau^{-2m+3}+
c_{\text{phas}}\tau^{2m-3}+ c_{\text{coupl}}\right]\tau^{-3}(t^{3D})^2$. There
is a minimum for $c_{\text{phon}}/c_{\text{phas}}=\tau^{4m-6}$. This
determines the Burgers vector with lowest energy for given values for the
material constants $c_{\text{phon}}$ and $c_{\text{phas}}$. However it has to
be noted that these are not necessarily identical for the $\Xi$- and the
$T$-approximants.

\begin{figure}
\begin{center}
\includegraphics[width=15cm]{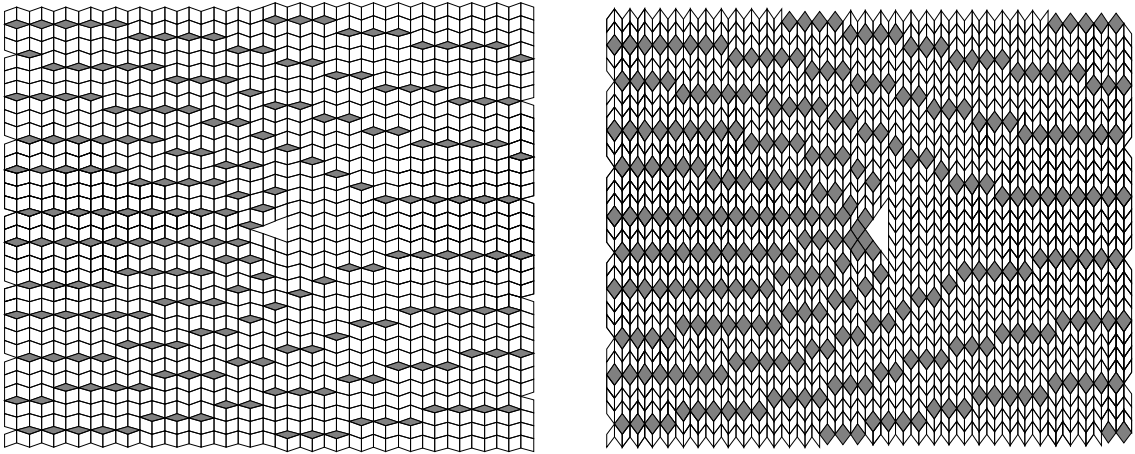}
\caption{Tilings of the $\xi'_{3}$- (left) and the $T_{3}$-phase (right) with
  $m=4$ metadislocations. $2F_{m}$ new phason planes are inserted from the
  left, ending at the triangular shaped dislocation core.
\label{fig:meta}}
\end{center}
\end{figure}

Tilings of the $\xi'_{3}$- and the $T_{3}$-phase with $m=4$ dislocations
have been calculated (Fig.~\ref{fig:meta}). They show large rearrangements
of the tiles due to the phasonic component $\ve{b}^{\perp}$ and negligible
deformations of the tiles due to the smaller phononic component
$\ve{b}^{\parallel}$. The dislocations are also dislocations in the
metastructure of the phason planes. Therefore \cite{itapdb:Klein1999}, who
discovered these dislocations in HRTEM images of the $\xi'_{2}$-phase, named
them metadislocations. The fact, that the experimentally most often observed
metadislocations are those with $m=4$ suggests
$c_{\text{phon}}/c_{\text{phas}}=\tau^{10}\simeq 123$. We do not know
of Burgers vector determinations or observations of metadislocations in the
$T$-approximants, but dislocations with the Burgers vectors (\ref{eq:burgers})
are also the ones most often observed in the i-phase
\cite[]{itapdb:Rosenfeld1995a}.

\section{Discussion and conclusion}

In the $\xi'_{n}$- and the $T_{n}$-phases the phasonic degree of freedom is
related to the movement of the phason planes. Since there are no phason planes
in the $\xi$-, $\xi'$-, $R$-, and $T$-phase, the phasonic degree of freedom
cannot be excited locally. However metadislocations can exist in the $\xi'$-
and the $T$-phase, but not in the $\xi$- and the $R$-phase. It can be shown,
that there is no consistent way to introduce dislocations with phasonic
components in the latter phases.

A motion of the metadislocation (like the motion of any dislocation in a
quasicrystal or large unit cell approximant) is necessarily accompanied by
diffusion in the form of tile rearrangements. The motion is possible by climb
in direction of the phason planes or by glide perpendicular to them. During
the climb motion new phason planes are created (or dissolved) behind the
dislocation core. A large number of metadislocations moving through the
$\xi'$- or $T$-phase could even lead to a phase transformation to the
$\xi'_{n}$- or $T_{n}$-phases making the phasonic degree of freedom
continously excitable. This is affirmed by HRTEM images of phase boundaries
between the $\xi'$- and the $\xi'_{2}$-phase formed by metadislcations
\cite[]{itapdb:Heggen2005}.

On the other side, glide motion does not change the number of phason
planes. At least in the $\xi'$- and the $T$-phase glide motion seems
unprobable, since the phason planes running out of the dislocation core would
have to be dragged along, while climb motion only needs a reconstruction of
the tiling near the dislocation core. (Similar arguments leading to the same
conclusion, as well as newer experimental work are presented in
\cite{itapdb:Feuerbacher2005}.) We have to note, that there are no direct
observations of metadislocation motion in approximants yet, although in the
i-phase dislocations with identical Burgers vectors have been shown by in-situ
observations to move by climb \cite[]{itapdb:Mompiou2004}.

\end{document}